\documentclass[journal]{IEEEtran}
\ifCLASSINFOpdf
\else
\fi
\usepackage{graphicx}
\graphicspath{{/Figures/}}
\DeclareGraphicsExtensions{.eps,.pdf,.png,.jpg}
\usepackage{subfig}
\usepackage{amsmath}
\usepackage{algorithm}
\usepackage{algorithmic}
\usepackage{cite}
\usepackage{amsthm}
\usepackage{amssymb}
\usepackage{color}
\usepackage{multicol}
\usepackage{bm}
\usepackage{tabularx}
\usepackage{booktabs}
\usepackage{balance}
\usepackage[top=0.75in, bottom=1in, left=0.635in, right=0.635in]{geometry}

\theoremstyle{definition}

\theoremstyle{remark}

\theoremstyle{proposition}

\begin{document}
\title{Software-defined and Virtualized Cellular Networks with M2M Communications}
\author{\IEEEauthorblockN{Meng Li\IEEEauthorrefmark{1}\IEEEauthorrefmark{2}, F.~Richard~Yu\IEEEauthorrefmark{3}, Pengbo Si\IEEEauthorrefmark{1}\IEEEauthorrefmark{2}, Enchang Sun\IEEEauthorrefmark{1}\IEEEauthorrefmark{2}, and Yanhua Zhang\IEEEauthorrefmark{1}\IEEEauthorrefmark{2}}\\
\thanks{This paper is supported by Natural Science Foundation of China under Grants No. 61372089 and No. 61571021.}
\IEEEauthorblockA{\IEEEauthorrefmark{1}Beijing Advanced Innovation Center for Future Internet Tech., Beijing Univ. of Tech., Beijing, P.R. China}\\
\IEEEauthorblockA{\IEEEauthorrefmark{2}College of Electronic Info. and Control Eng., Beijing Univ. of Techn., Beijing, P.R. China}\\
\IEEEauthorblockA{\IEEEauthorrefmark{3}Depart. of Systems and Computer Eng., Carleton Univ., Ottawa, ON, Canada}\\
Email: limeng$0720$@emails.bjut.edu.cn, richard.yu@carleton.ca, \{sipengbo, ecsun, zhangyh\}@bjut.edu.cn
}

\maketitle
\thispagestyle{empty}
\pagestyle{empty}
% As a general rule, do not put math, special symbols or citations in the abstract or keywords.
\begin{abstract}
Machine-to-machine (M2M) communications have attracted great attention from both academia and industry. In this paper, with recent advances in wireless network virtualization and software-defined networking (SDN), we propose a novel framework for M2M communications in software-defined cellular networks with wireless network virtualization. In the proposed framework, according to different functions and quality of service (QoS) requirements of machine-type communication devices (MTCDs), a hypervisor enables the virtualization of the physical M2M network, which is abstracted and sliced into multiple virtual M2M networks. Moreover, we formulate a decision-theoretic approach to optimize the random access process of M2M communications. In addition, we develop a feedback and control loop to dynamically adjust the number of resource blocks (RBs) that are used in the random access phase in a virtual M2M network by the SDN controller. Extensive simulation results with different system parameters are presented to show the performance of the proposed scheme.
\end{abstract}

% Note that keywords are not normally used for peer review papers.

% \fi
\IEEEpeerreviewmaketitle

\section{Introduction}
Machine-to-machine (M2M) communications, also named as machine-type communications (MTC), have attracted great attention in both academia and industry~\cite{LA14}. Unlike traditional human-to-human (H2H) communications (e.g., voice, messages, and video streaming) \cite{MYL04, YL01, LYH10, YK07, XYJL12, ATV12, YWL06_MONET,LY15, WYS10, GYJ10, BYC12, XYJ12, YTH09, YHT10, LYJ10, YKL06,YZX11, LYJ15,GYJ11,BY14,LYL09,YYG15,BYY15,ZYN12_JSAC,ZYL10,WTY14,BYL11_Online,BY13}, M2M communications have two main distinct characteristics: one is the large and rapid increasing number of MTCDs in the network (e.g., smart power grids, intelligent transportation, e-health, surveillance)~\cite{SR15}, the other is the data transmission in each time slot, which is mostly small-sized but the frequency of their making data connections is higher than traditional communication devices due to their specific roles and functions~\cite{LK11}. Based on these characteristics, how to support more MTCDs simultaneously connecting and accessing to the cellular network is an important and inevitable issue~\cite{IT14}. The authors of~\cite{OH15} propose a concept of random access efficiency, and formulate an optimization problem to maximize the random access efficiency with the delay constraint, according to the number of random access opportunities (RAOs) and MTCDs. In~\cite{HH13}, the authors introduce several random access (RA) overload control mechanisms to avoid collisions. The authors of~\cite{JKK14} investigate a scheme that provides additional preambles by spatially partitioning a cell coverage into multiple group regions and reducing cyclic shift size in RA preambles.

Although some excellent works have been done on M2M communications, most existing researches focus on preamble collision avoidance mechanisms.  However, most existing researches focus on preamble collision avoidance mechanisms. However, in practical networks, the MTCDs may fail to access the network if there is no enough radio resource allocated to the RA process~\cite{ZK15}. Furthermore, only one class of MTCDs are considered in most existing works. However, in practical networks, different MTCDs have different quality of service (QoS) requirements, and they should be treated differently in M2M communications.\

In this paper, with recent advances in \emph{wireless network virtualization}~\cite{LY15,LYY16} and \emph{software-defined networking} (SDN)~\cite{KR15}, we propose a novel framework for M2M communications in software-defined cellular networks with wireless network virtualization. In the proposed framework, according to different functions and classes of MTCDs, a hypervisor enables the virtualization of the physical M2M network, which is abstracted and sliced into multiple virtual M2M networks. Meanwhile, we formulate the random access process in M2M communications as a partially observable Markov decision process (POMDP). Moreover, we develop a feedback and control loop to dynamically adjust the number of resource blocks (RBs) that are used in the random access phase in a virtual M2M network. According to difference between the obtained and the desired transmission rate in virtual networks, the number of RBs are dynamically adjusted and allocated through the control loop by the SDN controller. \

The rest of this article is organized as follows. System model is presented in Section \ref{sec:Systemmodel}. In Section \ref{sec:Algorithm1}, we present an optimization algorithm for the random access process via POMDP formulation. Then resource allocation based on the feedback and control loop is formulated in Section \ref{sec:Algorithm2}. Section \ref{sec:Simulation} discusses the simulation results. Finally, we conclude this work in Section \ref{sec:Conclusion} with future works.\

\section{System Model}\label{sec:Systemmodel}
In this section, we develop the system model for the software-defined cellular network with M2M communications and network virtualization, the key components of the proposed framework are described as follows.

\subsection{Physical Resource Layer}
\begin{figure}[!t]
\centering
\includegraphics[width=2.8in]{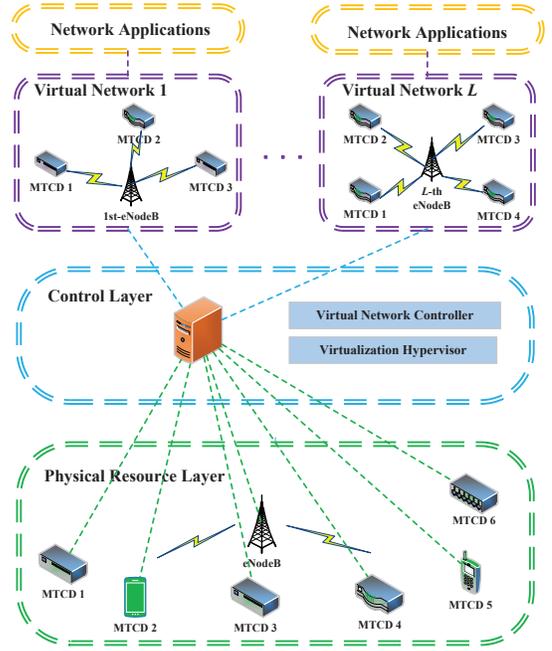}
 %where an .eps filename suffix will be assumed under latex,
% and a .pdf suffix will be assumed for pdflatex; or what has been declared
% via \DeclareGraphicsExtensions.
\vspace{0.5cm}
\caption{The architecture of a software-defined cellular network with M2M communications and wireless network virtualization.}
\label{fig:model}
\end{figure}

\
As shown in Fig. \ref{fig:model}, we consider the single-cell scenario with $N$ MTCDs and one eNodeB in the physical network. The time point that the MTCDs access the eNodeB is $t_{0}, t_{1},\ldots, t_{k},\ldots, t_{K-1}$, where $K$ is the total number of time slots, $1\leq k\leq K-1$, and each time slot is equal. It represents as $t_{k}-t_{k-1}=\delta t_k$, where $\delta t_k$ is the duration of a time slot. A time period includes the $K$ time slots, from time point $t_{0}$ to $t_{K-1}$, each time period is represented as $T_1,T_2,\ldots,T_{y},\ldots,T_{Y}$. Meanwhile, RBs will be offered by the eNodeB when the MTCDs attempt to access the eNodeB. We assume that the total number of RBs is $R_{total}$. The number of RBs used in the control access phase is $R$, while that used in the data transmission phase is $R^{'}$. They satisfy that $R+R^{'}=R_{total}$. Considering the RBs for the access phase, $r$ represents the $r$-th RB, where $1\leq r\leq R$. The state of each RB in one time slot can be described as idle or busy. We use the set $\bm{s_r}$ to represent the state of the $r$-th RB, and $\bm{s_{r}}=\{0,1\}$, where $0$ stands for the RB is idle while $1$ stands for the RB is busy in this time slot.\

We assume that each RB can offer different transmission rates for MTCDs. After the $n$-th MTCD has accessed to the $r$-th RB, we define $C_{n,r}(k)$ as the transmission rate achieved by MTCD in time slot $\delta t_{k}$, and it can be calculated as\
\begin{eqnarray}
&& C_{n,r}(k)= \nonumber \\
&& \left\{
\begin{array}{lcl}
B_{n,r} \log_2 \left\{ 1+\frac{P_{r}h_{n,r}}{\sigma^{2}}\right\}, \text{if}\ s_{r}=0,\\
B_{n,r} \log_2 \left\{1+\frac{P_{r}h_{n,r}}{\sum\limits_{n^{'}\neq n,n^{'}\in N} P_{r}h_{n^{'},r}+\sigma^{2}}\right\}, \text{if}\ s_{r}=1,\\
\end{array}
\right.
\end{eqnarray}
where $B_{n,r}$ represents the bandwidth offered by the $r$-th RB, $P_{r}$ represents the transmit power consumed by the $r$-th RB, $h_{n,r}$ ($h_{n^{'},r}$) is the channel gain when the $n$-th ($n^{'}$-th) MTCD accesses to the RB, which follows Gaussian distribution with zero mean and unit variance, and $\sigma^{2}$ is the system noise power.\

\subsection{Control Layer}\
In the proposed framework, the controller is set in this layer, which includes the hypervisor and SDN controller. The hypervisor is an important component in wireless network virtualization. In general, the hypervisor can be implemented at the physical eNodeB, and it provides functions to connect physical resource and virtual eNodeB~\cite{LY15}. Moreover, the hypervisor takes the responsibility of virtualizing the physical eNodeB into a number of virtual eNodeBs~\cite{LYZ15}. Besides, the hypervisor is also responsible for scheduling the air interface resources. As mentioned above, the SDN controller also plays an essential role in the proposed framework, and the network resources can be allocated dynamically by the SDN controller~\cite{CYL15,WL16}.\

In this paper, the physical network is abstracted and sliced into virtual networks by the hypervisor. Meanwhile, a feedback control loop is proposed and designed in the control layer, then all of RBs offered by the eNodeB can be allocated dynamically to each virtual network by the SDN controller. According to different functions of different virtual networks, the SDN controller can adjust the number of allocated RBs to optimize and improve the performance of networks~\cite{BOS14}. By this means, in the virtual network with M2M communications, the SDN controller will offer an efficient approach to allocate RBs for M2M communications.\

\subsection{Virtual Network Layer}\
As shown in Fig. \ref{fig:model}, according to different QoS requirements, the physical network will be virtualized to multiple virtual networks by hypervisor. The hypervisor takes the responsibility of mapping the physical network with M2M communications into $L$ virtual networks. For the $l$-th $(1\leq l\leq L)$ virtual network, it includes $N_l (1\leq N_l\leq N)$ MTCDs, which have the same or similar function. Meanwhile, in the $l$-th virtual network, the virtual eNodeB can offer all RBs to control access and data transmission in the initial time slot. The numbers of RBs used for control access and data transmission are $R_l (1\leq R_l\leq R)$ and $R_{l}^{'} (1\leq R_l^{'}\leq R^{'})$, respectively. In addition, the SDN controller can dynamically allocate the physical resources for each virtual network, and it also can provide and manage specific services to MTCDs.\

The obtained transmission rate of each virtual network can be denoted as $C_{1}, C_{2}, \dots, C_{l}, \dots, C_{L}$, where $C_{1}$ represents the obtained transmission rate in the highest level virtual network and $C_{L}$ represents the obtained transmission rate in the lowest level virtual network. To provide the proportional average transmission rate differentiation, the average transmission rate of the $L$ levels should be related by the expression
\begin{eqnarray}
C_1:C_2:\ldots:C_l:\ldots:C_L\approx x_1:x_2:\ldots:x_l:\ldots:x_L,
\end{eqnarray}
where $x_l$ represents a constant weighting factor for level requirement of the $l$-th virtual network. Obviously, it satisfies that $x_{1}\geq x_{2}\geq\ldots x_{l}\geq\ldots\geq x_{L}$. For the $l$-th virtual network, the obtained average transmission rate $C_l$ can be calculated as
\begin{eqnarray}
C_l=\frac{\sum\limits_{r=1}^{R_l}\sum\limits_{k=1}^{K}{{C_{l,n,r}(k)}\delta t_k}}{T_y}.
\end{eqnarray}

Then, the ratio of obtained and desired transmission rate in the $l$-th virtual network can be denoted as
\begin{eqnarray}\label{accessrate1}
\xi_l=\frac{C_l}{C_1+C_2+\ldots+C_l+\ldots+C_L},
\end{eqnarray}
\begin{eqnarray}\label{accessrate2}
\xi^{'}_l=\frac{x_l}{x_1+x_2+\ldots+x_l+\ldots+x_L},
\end{eqnarray}
where $\xi_l$ denotes the ratio of obtained transmission rate, and $\xi^{'}_l$ denotes the ratio of desired transmission rate. Therefore, the gap between the ratio of desired and obtained transmission rate can be written as $e_l=\xi^{'}_l-\xi_l$. Thus, $e_l$ is used by the SDN controller to decide the RBs adjustment and allocation in the access phase. According to Eqs. (\ref{accessrate1}) and (\ref{accessrate2}), both $\xi_l$ and $\xi^{'}_l$ are used as the performance metrics of the feedback control loop.\

\section{Optimization of Random Access via POMDP}\label{sec:Algorithm1}
In this section, we develop a decision-theoretic approach via POMDP to optimize the random access process. Then, each tuple of POMDP is described in detail, followed by the reward and optimization objective.\

\subsection{POMDP Formulation}
Since the state of RBs cannot be directly observed by MTCDs, the problem of random access can be formulated as a POMDP optimization problem~\cite{LM14}. For simplicity, the POMDP formulation is discussed by taking the $l$-th virtual network as an example.\

\emph{1) Action Space} \

 Let $\bm{A}$ represent the set of all available actions, and the action that can be taken by this MTCD in time slot $\delta t_k$ can be defined as
\begin{equation}
\begin{aligned}
a(k) \in \{0 (no\ access),{RB_1},{RB_2},\dots,{RB_r},\dots,{RB_{R_{l}}}\}.
\end{aligned}
\end{equation}

In set $\bm{A}$, $0$ represents that the MTCD will not access the eNodeB and select sleeping mode,  $RB_r$ represents that the MTCD will select the $r$-th RB to access to the eNodeB.\

\emph{2) State Space and Transition Probability}\

In the M2M communication network, the system state space $\bm{S}$ is the set of all RB states, and the state in time point $t_k$ can be denoted as $s(k)=[{s_{1}(k)}{s_{2}(k)}\dots{s_{r}(k)}\dots{s_{R_l}(k)}]$, where $s(k)\in\bm{S}$. Note that, the state of the $r$-th RB can be defined as
 \begin{equation}
\begin{aligned}
{s_{r}(k)} \in \{0 (idle),1(busy)\}.
\end{aligned}
\end{equation}

Assume that each RB state is discretized, and the number of busy RBs in each time slot can be modelled as a random process with Possion distribution. We consider that $p_{i,j}$ is the transition probability of the RB state from state $i$ to state $j$ and can be expressed as
 \begin{equation}\label{stateprobability}
\begin{aligned}
p_{ i,j}=Prob.\{s_r(k+1)=j\mid s_r(k)=i\}.
\end{aligned}
\end{equation}

\emph{3) Observation Space}\

Since it is difficult to acquire the full knowledge of each RB state, the MTCD needs to observe the  RB state based on the state transition and optimal action taken in this time slot~\cite{LM14}. Let $\theta_{r}(k)$ denote the observation state of the $r$-th RB in time slot $\delta t_k$, where $1\leq r\leq R_l$. $\theta_{r}(k)$ can be identified as
\begin{equation}
\begin{aligned}
{\theta_{r}(k)} \in \{0 (idle),1(busy)\}.
\end{aligned}
\end{equation}
Then in time slot $\delta t_k $, the observation state can be written as ${\theta(k)}=[\theta_{1}(k)\theta_{2}(k)\dots\theta_{r}(k)\dots\theta_{R_l}(k)]$, where $\theta(k)\in\Theta$, and $\Theta$ is the set of all observation states.

As the $r$-th RB state transits from $s_r(k)$ to $s_r(k +1)$ under action $a(k)$, an observation state $\theta_r(k)$ is generated with the conditional probability $b^{a(k)}_{s_r(k+1),\theta_r(k)}=Pr\{\theta_r(k) \mid s_r(k+1), a(k)\}$. Hence, the conditional probability of observation can be denoted as
\begin{eqnarray}
%\begin{displaymath}
b^{a(k)}_{s_r(k+1),\theta_r(k)}=\left\{
\begin{array}{lll}
\epsilon, &\text {if}~a(k)=RB_r,~\theta_r(k)=0,\\
1-\epsilon, &\text {if}~a(k)=RB_r,~\theta_r(k)=1,\\
\varphi, &\text {if}~a(k)=0,~\theta_r(k)=0,\\
1-\varphi, &\text {if}~a(k)=0,~\theta_r(k)=1,\\
\end{array}
\right.
\end{eqnarray}
%\end{displaymath}
where $\epsilon$ and $\varphi$ are the probability of false observation, i.e., mistaking the busy state for idle state. For the sake of simplicity, in the proposed scheme, we assume that $\epsilon=\varphi$.

\emph{4) Information State}\

Let $\pi(k)=\{\pi_1^{k}, \pi_2^{k},\dots,\pi_{i}^{k},\dots, \pi_{s_{R_l}}^{k}\}$ denote the information space, where $\pi_{i}^{k}\in[0,1]$ is the conditional probability (given decision and observation history) that the RB state is in $i$ at the beginning of time slot $\delta t_k$ prior to state transition.\

The information state can be easily updated after each state transition to incorporate additional step information into history, and it is updated by using Bayes' rule at the end of each time slot~\cite{LM14,SS73}, it can be represented as follows,
\begin{equation}
\pi_{s_r(k+1)}^{k+1}=\frac{\sum_{s_r(k)}\pi_{s_r(k)}^{k}p_{s_r(k), s_r(k+1)}b^{a(k)}_{s_r(k+1),\theta_r(k)}}{\sum_{s_r(k), s_r(k+1)} \pi_{s_r(k)}^{k}p_{s_r(k), s_r(k+1)}b^{a(k)}_{s_r(k+1),\theta_r(k)}}.
\end{equation}

\emph{5) Reward and Objective}\

By regarding the transmission rate as a reward, the maximum transmission rate offered by RB can be used for performance evaluation. Since each system state is decided by all $R_l$ RBs states, the maximum value of the transmission rate offered by RB will be taken as the reward. Hence, for each system state, the corresponding transmission rate can be denoted as
\begin{equation}
C_{l,n}(k)=\max\limits_{{r}\in[1,R_l]}\{C_{l,n,1}(k),\dots, C_{l,n,r}(k),\dots, C_{l,n,R_l}(k)\},
\end{equation}
where $C_{l,n,r}(k)$ is the transmission rate offered by the $r$-th RB in time slot $\delta t_k$.

Then the optimization objective is to maximize the transmission rate that can be achieved by MTCDs. Therefore, the system reward in the proposed scheme within time slot $\delta t_k$ is originally defined as
\begin{eqnarray}
%\begin{displaymath}
Re_{l,n}(k)=\left\{
\begin{array}{lll}
0, &\text {if there is no sensing},\\
C_{l,n}(k), &\text {otherwise},\\
\end{array}
\right.
\end{eqnarray}
and the total discounted reward $Re_{l,n}$ is
\begin{eqnarray}
Re_{l,n}=\sum\limits^{K-1}_{k=0}\beta^{K-k-1}Re_{l,n}(k),
\end{eqnarray}
where $\beta\in[0,1]$ is the discount factor.\

The optimal policy $\bm{U}$ in this paper is represented as the set of behaviour $a(k)$, $0\leq k\leq K-1$, which maximises the expected long-term total discounted reward $Re_{l,n}$ during a time period. Hence, the optimal policy is represented as
\begin{eqnarray}
\bm{U}=\{a(k)\}=\arg\max\limits_{a(k)\in \mathcal{A}}E[Re_{l,n}].
\end{eqnarray}

\subsection{Solving the POMDP Problem}
Let $J_{k}(\pi(k))$ be the maximum expected reward that can be obtained from time slot $\delta t_{k}$, given the information state $\pi(k)$ at the beginning of time slot $\delta t_{k}$. Assuming that the MTCD that attempts to access the RB makes action $a(k)$ and observes state $\theta_r(k)$, the reward can be accumulated starting from time slot $\delta t_{k}$. It should be noticed that the reward includes two parts~\cite{LY10}: one is the immediate reward $Re_{l,n}$, the other is the maximum expected future reward $J_{k+1}(\pi(k+1))$ starting from time slot $\delta t_{k+1}$, given the information state $\pi(k+1)$. As a result, the optimal policy of random access can be calculated as
\begin{equation}\label{totalreward}
\begin{split}
J_{k}(\pi(k))=\max\limits_{a(k)\in\bm{A}}\sum\limits_{s_r(k)\in \bm{S}}\sum\limits_{s_r(k+1)\in \bm{S}}\pi_{{s}_r(k)}^{k}p(s_r(k),s_r(k+1))\\ \sum\limits_{s_r(k+1)\in \bm{S}}b^{a(k)}_{s_r(k+1),\theta_r(k)}[Re_{l,n}(k)+J_{k+1}(\pi(k+1))],\\ \forall 1\leq k\leq K-1.
\end{split}
\end{equation}

\section{Resource Allocation via Feedback and Control}\label{sec:Algorithm2}
In this section, we will present a strategy of feedback and control to allocate RBs that are used in the random access phase by SDN controller. After that, a detailed design method of the control loop will be given, and a novel approach for RBs allocation and adjustment with M2M communications will be proposed.\

\subsection{Resource Allocation with SDN Controller}
\begin{figure}[!t]
\centering
\includegraphics[width=3.5in]{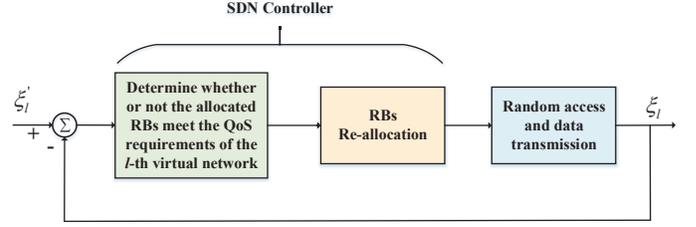}
 %where an .eps filename suffix will be assumed under latex,
% and a .pdf suffix will be assumed for pdflatex; or what has been declared
% via \DeclareGraphicsExtensions.
\caption{The feedback and control loop for RBs allocation.}
\label{fig:allocation1}
\end{figure}

After a time period $T_{y}$, if the obtained average transmission rate cannot reach the desired one, a virtual network needs a feedback mechanism to adjust RBs allocation in the access phase based on the gap of ratio between the obtained and desired average transmission rate. The proposed feedback mechanism is depicted in Fig. \ref{fig:allocation1}.\

\begin{figure}[!t]
\centering
\includegraphics[width=3.5in]{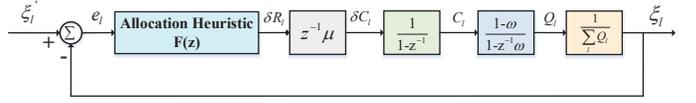}
 %where an .eps filename suffix will be assumed under latex,
% and a .pdf suffix will be assumed for pdflatex; or what has been declared
% via \DeclareGraphicsExtensions.
\caption{$z$-transform used in the feedback and control loop with M2M communications.}
\label{fig:allocation2}
\end{figure}

In each virtual network, the number of RBs that are assigned by the virtual eNodeB is fixed in the access phase. Based on the ratio of obtained and desired transmission rate in the $l$-th virtual network, which is calculated by $\xi_l$ and $\xi_{l}^{'}$, the RB allocation algorithms through the control loop can be developed. With the proposed algorithm, the objective converts to adjust the RB allocation between random access phase and data transmission phase, or among virtual networks by the SDN controller. Moreover, let the gap of ratio between the obtained and desired transmission rate after a time period $T_{y}$ be $e_l[T_{y}]$. In order to compute the reassignment number of RBs $\delta R_l[T_{y}]$, the SDN controller will utilize a linear function $f(e_l)$ and compute $\delta R_l[T_{y}]$ as follows
\begin{eqnarray}\label{numbergap1}
\forall l: \delta R_l[T_{y}]=f(e_l[T_{y}]),
\end{eqnarray}
and the number of RBs in time period $T_{y}$ is adjusted as
\begin{eqnarray}\label{numbergap2}
\forall l: R_l[T_{y}]=R_l[T_{y-1}]+\delta R_l[T_{y}].
\end{eqnarray}

According to Eqs. (\ref{numbergap1}) and (\ref{numbergap2}), the allocation strategy is concluded as: if the correction $\delta R_l[T_{y}]$ is positive, the number of RBs allocated to the $l$-th virtual network in the access phase is increased by $\mid\delta R_l[T_{y}]\mid$; otherwise, it will be decreased by that number.

\subsection{Feedback and Control Loop Design}
In this subsection, a control loop-based model is proposed in order to design  function $f(e_l)$. In essence, an approximate linear model is alternative to simplify the design of the feedback control mechanism, since the nonlinear relationship between the adjustment number of RBs and the gap rate~\cite{LA04}.\

Due to the linear allocation behavior, the relationship between the variation of average transmission rate and the adjustment number of RBs is approximatively proportional and can be described as
\begin{eqnarray}
\delta C_l[T_{y}]\approx\mu\delta R_l[T_{y-1}],
\end{eqnarray}
where $\mu$ is a proportionality coefficient. Then the obtained transmission rate and the variation of transmission rate should satisfy
\begin{eqnarray}
C_l[T_y]=C_l[T_{y-1}]+\delta C_l[T_y].
\end{eqnarray}

Considering Eq. (\ref{accessrate1}), it is worth noting that the obtained average transmission rate $C_l[T_y]$ might have a large standard deviation, compared with the expected value except that the time period is sufficiency large. In order to solve this problem, a low pass filter will be applied in the feedback loop. By letting $Q_l[T_y]$ be the output of $C_l[T_y]$ through the smooth filter, it follows that
\begin{eqnarray}
Q_l[T_y]=\omega Q_l[T_{y-1}]+(1-\omega)C_l[T_y],
\end{eqnarray}
where $\omega$ is a factor and satisfies that $0<\omega<1$.\

As can be seen in Fig. \ref{fig:allocation2}, the control loop shows the process and relationship in the $z$-transform. The function $f$ with respect to the RB number adjustment through $z$-transform can be expressed as $F(z)$.\

According to Fig. \ref{fig:allocation2}, $\xi_l$ can be denoted as
\begin{eqnarray}
\xi_l[T_y]=e_lF(z)G(z),
\end{eqnarray}
where
\begin{eqnarray}\label{ztransform1}
G(z)=\frac{z^{-1}\mu(1-\omega)}{(1-z^{-1})(1-z^{-1}\omega)\sum\limits_{l=1}^{L}Q_l}.
\end{eqnarray}
Then, by substituting for $e_l$ and using simple algebraic manipulation, the relationship  between the obtained and desired transmission rate can be represented as
\begin{eqnarray}\label{ztransform2}
\xi_l=\frac{F(z)G(z)}{1+F(z)G(z)}\xi^{'}_l.
\end{eqnarray}

In order to design the RB number allocation in accord with desired behavior of the closed loop, $\xi_l[T_y]$ should follow $\xi^{'}_l[T_y]$ within one time period. In the $z$-transform, the corresponding condition according to Eqs. (\ref{ztransform1}) and (\ref{ztransform2}) can be represented as
\begin{eqnarray}\label{ztransform3}
\frac{F(z)G(z)}{1+F(z)G(z)}=z^{-1}.
\end{eqnarray}
Meanwhile, substituting for $G(z)$ into Eq. (\ref{ztransform3}), $F(z)$ is represented as
\begin{eqnarray}\label{ztransform4}
F(z)=\frac{(1-z^{-1}\omega)\sum\limits_{l=1}^{L}Q_l}{\mu(1-\omega)}.
\end{eqnarray}

At last, the SDN controller will adjust the number of RBs based on $\delta R_l[T_y]$ from the data transmission phase or other virtual networks to the access phase. According to the $z$-inverse transform in Eq. (\ref{ztransform4}), $\delta R_l[T_y]$ can be calculated as
\begin{eqnarray}
\delta R_l[T_y]=f(e_l)=\frac{\sum\limits_{l=1}^{L}Q_l}{\mu(1-\omega)}(e_l[T_y]-\omega e_l[T_{y-1}]).
\end{eqnarray}

\section{Simulation Results and Discussions}\label{sec:Simulation}
In this section, simulation results are presented to show the performance of the proposed scheme with the random access optimization modeled by POMDP and RBs allocation realized by the control loop.\

We consider a single-cell scenario with one eNodeB and 50 MTCDs. Meanwhile, 25 RBs can be offered by eNodeB.  We assume that the physical network is sliced into 5 virtual networks according to the function of MTCDs. For each virtual network, it consists of one virtual eNodeB and several MTCDs. MTCDs will be distributed uniformly, with $N_1=30$ and $N_l=5$ ($l=2, 3, \dots, 5$).  In the initial time slot, each virtual eNodeB will be allocated $5$ RBs. The probability that RB remains in the idle state, remains in the busy state, transit from busy to idle state and transit from busy to idle state is set as $0.9$, $0.05$, $0.95$ and $0.1$, respectively. The probability of false observation ranges from $0.1$ to $0.8$. Additionally, the available transmission bandwidth in the first and the fifth virtual network is $10$ MHz and $5$ MHz, respectively. The transmit power is $20$ dBm in both virtual networks. Channel gains follow Gaussian distribution with zero mean and unit variance. Moreover, the weighting factor is set as $x_1:x_5=3:1$. In addition, the factor $\omega$ is $0.8$, and the proportionality coefficient $\mu$ is $2$.\

\begin{figure}[!t]
\centering
\includegraphics[width=3.2in]{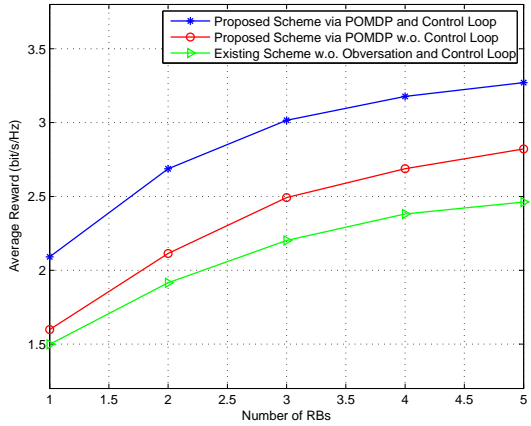}
 %where an .eps filename suffix will be assumed under latex,
% and a .pdf suffix will be assumed for pdflatex; or what has been declared
% via \DeclareGraphicsExtensions.
\caption{Average reward with different numbers of RBs in the heterogeneous traffic scenario.}
\label{fig:RewardRB}
\end{figure}

Fig. \ref{fig:RewardRB} compares the reward with different numbers of RBs in heterogeneous traffic scenario. In detail, with only one RB, there is little difference between the proposed scheme via POMDP without control loop and the existing scheme, since there is no decision flexibility. However, for the proposed scheme via POMDP and control loop, the transmission performance is improved significantly, the average reward in the proposed scheme via POMDP and control loop is much higher than other schemes without the
control loop. The reason is that the control loop can adjust the number of RBs to meet the network requirements. With the increasing number of RBs, the proposed scheme via POMDP and control loop is more prominent than other schemes, since more RBs can be offered and more selections can be made by the POMDP optimization.\

\begin{figure}[!t]
\centering
\includegraphics[width=3.2in]{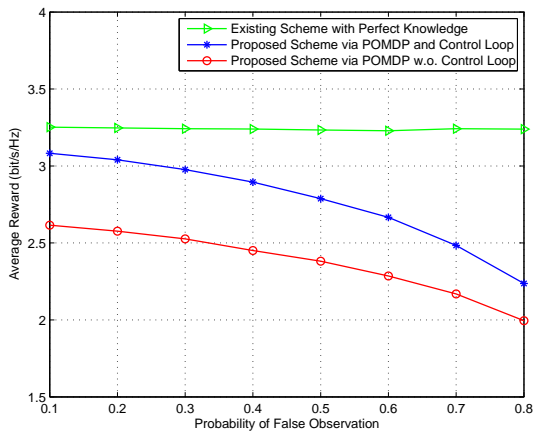}
 %where an .eps filename suffix will be assumed under latex,
% and a .pdf suffix will be assumed for pdflatex; or what has been declared
% via \DeclareGraphicsExtensions.
\caption{Average reward with different probabilities of false observation in the heterogeneous traffic scenario.}
\label{fig:Rewardobserve}
\end{figure}

Fig. \ref{fig:Rewardobserve} depicts the variation of the average reward with different probabilities of false observation in heterogeneous traffic scenario. It can be easily seen that the average reward in the proposed scheme degrades with the increasing probability of false observation. When $\epsilon=\varphi=0.1$, the proposed scheme with the control loop will be close to the existing scheme with perfect knowledge. However, if $\epsilon=\varphi=0.8$, the performance in the proposed scheme degrades obviously. The reason is that MTCDs have to give up or falsely select RBs with poor performance to access when the probability of false observation reaches high value resulting in a lower average reward.\

\section{Conclusions and Future Work}\label{sec:Conclusion}
In this paper, we proposed a novel framework for M2M communications in software-defined cellular networks with wireless network virtualization. In the proposed framework, we formulated the random access process as a POMDP, by which MTCDs can select proper RB to achieve the maximum transmission rate. In addition, a feedback and control loop was developed to adjust and allocate RBs by the SDN controller after each time period. With virtual resource allocation in each virtual network, the obtained transmission rate approaches the desired one. Simulation results demonstrated that, with the proposed framework, the number of RBs can be dynamically adjusted according to the gap between the ratio of the obtained and the desired transmission rate in each virtual network. Future work is in progress to consider energy consumption and cooperative communications in our framework.\

\balance
\bibliographystyle{IEEEtran}
\bibliography{limengreference,D:/CA/Papers/Ref}

\begin{thebibliography}{10}
\providecommand{\url}[1]{#1}
\csname url@samestyle\endcsname
\providecommand{\newblock}{\relax}
\providecommand{\bibinfo}[2]{#2}
\providecommand{\BIBentrySTDinterwordspacing}{\spaceskip=0pt\relax}
\providecommand{\BIBentryALTinterwordstretchfactor}{4}
\providecommand{\BIBentryALTinterwordspacing}{\spaceskip=\fontdimen2\font plus
\BIBentryALTinterwordstretchfactor\fontdimen3\font minus
  \fontdimen4\font\relax}
\providecommand{\BIBforeignlanguage}[2]{{%
\expandafter\ifx\csname l@#1\endcsname\relax
\typeout{** WARNING: IEEEtran.bst: No hyphenation pattern has been}%
\typeout{** loaded for the language `#1'. Using the pattern for}%
\typeout{** the default language instead.}%
\else
\language=\csname l@#1\endcsname
\fi
#2}}
\providecommand{\BIBdecl}{\relax}
\BIBdecl

\bibitem{LA14}
{A.~Laya}, {L.~Alonso}, and {J.~Alonso-Zarate}, ``Is the random access channel
  of {LTE} and {LTE-A} suitable for {M2M} communications? {A} survey of
  alternatives,'' \emph{IEEE Commun. Surveys Tutorials}, vol.~16, no.~1, pp.
  4--16, Jan.~2014.

\bibitem{MYL04}
L.~Ma, F.~Yu, V.~C.~M. Leung, and T.~Randhawa, ``A new method to support
  {UMTS/WLAN} vertical handover using {SCTP},'' \emph{IEEE Wireless Commun.},
  vol.~11, no.~4, pp. 44--51, Aug. 2004.

\bibitem{YL01}
F.~Yu and V.~C.~M. Leung, ``Mobility-based predictive call admission control
  and bandwidth reservation in wireless cellular networks,'' in \emph{Proc.
  IEEE INFOCOM'01}, Anchorage, AK, Apr. 2001.

\bibitem{LYH10}
Z.~Li, F.~R. Yu, and M.~Huang, ``A distributed consensus-based cooperative
  spectrum sensing in cognitive radios,'' \emph{IEEE Trans. Veh. Tech.},
  vol.~59, no.~1, pp. 383--393, Jan. 2010.

\bibitem{YK07}
F.~Yu and V.~Krishnamurthy, ``Optimal joint session admission control in
  integrated {WLAN} and {CDMA} cellular networks with vertical handoff,''
  \emph{IEEE Trans. Mobile Computing}, vol.~6, no.~1, pp. 126--139, Jan. 2007.

\bibitem{XYJL12}
R.~Xie, F.~R. Yu, H.~Ji, and Y.~Li, ``Energy-efficient resource allocation for
  heterogeneous cognitive radio networks with femtocells,'' \emph{IEEE Trans.
  Wireless Commun.}, vol.~11, no.~11, pp. 3910 --3920, Nov. 2012.

\bibitem{ATV12}
A.~Attar, H.~Tang, A.~Vasilakos, F.~R. Yu, and V.~Leung, ``A survey of security
  challenges in cognitive radio networks: Solutions and future research
  directions,'' \emph{Proceedings of the IEEE}, vol. 100, no.~12, pp.
  3172--3186, 2012.

\bibitem{YWL06_MONET}
Y.~Fei, V.~W.~S. Wong, and V.~C.~M. Leung, ``Efficient {QoS} provisioning for
  adaptive multimedia in mobile communication networks by reinforcement
  learning,'' \emph{Mob. Netw. Appl.}, vol.~11, no.~1, pp. 101--110, Feb. 2006.

\bibitem{LY15}
{C.~Liang} and {F. R.~Yu}, ``Wireless network virtualization: {A} survey, some
  research issues and challenges,'' \emph{IEEE Commun. Surveys Tutorials},
  vol.~17, no.~1, pp. 358--380, Mar.~2015.

\bibitem{WYS10}
Y.~Wei, F.~R. Yu, and M.~Song, ``Distributed optimal relay selection in
  wireless cooperative networks with finite-state {Markov} channels,''
  \emph{IEEE Trans. Veh. Tech.}, vol.~59, no.~5, pp. 2149 --2158, June 2010.

\bibitem{GYJ10}
Q.~Guan, F.~R. Yu, S.~Jiang, and G.~Wei, ``Prediction-based topology control
  and routing in cognitive radio mobile ad hoc networks,'' \emph{IEEE Trans.
  Veh. Tech.}, vol.~59, no.~9, pp. 4443 --4452, Nov. 2010.

\bibitem{BYC12}
S.~Bu, F.~R. Yu, Y.~Cai, and P.~Liu, ``When the smart grid meets
  energy-efficient communications: Green wireless cellular networks powered by
  the smart grid,'' \emph{IEEE Trans. Wireless Commun.}, vol.~11, pp.
  3014--3024, Aug. 2012.

\bibitem{XYJ12}
R.~Xie, F.~R. Yu, and H.~Ji, ``Dynamic resource allocation for heterogeneous
  services in cognitive radio networks with imperfect channel sensing,''
  \emph{IEEE Trans. Veh. Tech.}, vol.~61, pp. 770--780, Feb. 2012.

\bibitem{YTH09}
F.~R. Yu, H.~Tang, M.~Huang, Z.~Li, and P.~C. Mason, ``Defense against spectrum
  sensing data falsification attacks in mobile ad hoc networks with cognitive
  radios,'' in \emph{Proc. IEEE Military Commun. Conf. (MILCOM)'09}, Oct. 2009.

\bibitem{YHT10}
F.~R. Yu, M.~Huang, and H.~Tang, ``Biologically inspired consensus-based
  spectrum sensing in mobile ad hoc networks with cognitive radios,''
  \emph{IEEE Network}, vol.~24, no.~3, pp. 26 --30, May 2010.

\bibitem{LYJ10}
C.~Luo, F.~R. Yu, H.~Ji, and V.~C.~M. Leung, ``Cross-layer design for {TCP}
  performance improvement in cognitive radio networks,'' \emph{IEEE Trans. Veh.
  Tech.}, vol.~59, no.~5, pp. 2485--2495, 2010.

\bibitem{YKL06}
F.~Yu, V.~Krishnamurthy, and V.~C.~M. Leung, ``Cross-layer optimal connection
  admission control for variable bit rate multimedia traffic in packet wireless
  {CDMA} networks,'' \emph{IEEE Trans.\ Signal Proc.}, vol.~54, no.~2, pp.
  542--555, Feb. 2006.

\bibitem{YZX11}
F.~R. Yu, P.~Zhang, W.~Xiao, and P.~Choudhury, ``Communication systems for grid
  integration of renewable energy resources,'' \emph{IEEE Network}, vol.~25,
  no.~5, pp. 22 --29, Sept. 2011.

\bibitem{LYJ15}
{G.~Liu}, {F. R.~Yu}, {H.~Ji}, {V. C.~Leung}, and {X.~Li}, ``In-band
  full-duplex relaying: A survey, some research issues and challenges,''
  \emph{IEEE Commun. Surveys Tutorials}, vol.~17, no.~2, pp. 500--524, Second
  quarter 2015.

\bibitem{GYJ11}
Q.~Guan, F.~R. Yu, S.~Jiang, and V.~Leung, ``Capacity-optimized topology
  control for {MANETs} with cooperative communications,'' \emph{IEEE Trans.\
  Wireless Commun.}, vol.~10, no.~7, pp. 2162 --2170, July 2011.

\bibitem{BY14}
S.~Bu and F.~R. Yu, ``Green cognitive mobile networks with small cells for
  multimedia communications in the smart grid environment,'' \emph{IEEE Trans.
  Veh. Tech.}, vol.~63, no.~5, pp. 2115--2126, June 2014.

\bibitem{LYL09}
J.~Liu, F.~R. Yu, C.-H. Lung, and H.~Tang, ``Optimal combined intrusion
  detection and biometric-based continuous authentication in high security
  mobile ad hoc networks,'' \emph{IEEE Trans. Wireless Commun.}, vol.~8, no.~2,
  pp. 806--815, 2009.

\bibitem{YYG15}
Q.~Yan, F.~R. Yu, Q.~Gong, and J.~Li, ``Software-defined networking ({SDN}) and
  distributed denial of service ({DDoS}) attacks in cloud computing
  environments: A survey, some research issues, and challenges,'' \emph{IEEE
  Commun. Survey and Tutorials}, vol.~18, no.~1, pp. 602--622, 2016.

\bibitem{BYY15}
S.~Bu, F.~R. Yu, and H.~Yanikomeroglu, ``Interference-aware energy-efficient
  resource allocation for heterogeneous networks with incomplete channel state
  information,'' \emph{IEEE Trans. Veh. Tech.}, vol.~64, no.~3, pp. 1036--1050,
  Mar. 2015.

\bibitem{ZYN12_JSAC}
L.~Zhu, F.~R. Yu, B.~Ning, and T.~Tang, ``Cross-layer handoff design in
  {MIMO}-enabled {WLANs} for communication-based train control ({CBTC})
  systems,'' \emph{IEEE J. Sel. Areas Commun.}, vol.~30, no.~4, pp. 719--728,
  May 2012.

\bibitem{ZYL10}
S.~Zhang, F.~R. Yu, and V.~Leung, ``Joint connection admission control and
  routing in {IEEE} 802.16-based mesh networks,'' \emph{IEEE Trans.\ Wireless
  Commun.}, vol.~9, no.~4, pp. 1370 --1379, Apr. 2010.

\bibitem{WTY14}
Z.~Wei, H.~Tang, F.~R. Yu, M.~Wang, and P.~Mason, ``Security enhancements for
  mobile ad hoc networks with trust management using uncertain reasoning,''
  \emph{IEEE Trans. Veh. Tech.}, vol.~63, no.~9, pp. 4647--4658, Nov. 2014.

\bibitem{BYL11_Online}
S.~Bu, F.~R. Yu, and P.~Liu, ``Dynamic pricing for demand-side management in
  the smart grid,'' in \emph{Proc. IEEE Online Conference on Green
  Communications (GreenCom)'11}, Sept. 2011, pp. 47--51.

\bibitem{BY13}
S.~Bu and F.~R. Yu, ``A game-theoretical scheme in the smart grid with
  demand-side management: Towards a smart cyber-physical power
  infrastructure,'' \emph{IEEE Trans. Emerging Topics in Computing}, vol.~1,
  no.~1, pp. 22--32, June 2013.

\bibitem{SR15}
{H.~Shariatmadari}, {R.~Ratasuk}, {S.~Iraji}, {A.~Laya}, {T.~Taleb},
  {R.~Jantti}, and {A.~Ghosh}, ``Machine-type communications: current status
  and future perspectives toward 5{G} systems,'' \emph{IEEE Comm. Mag.},
  vol.~53, no.~9, pp. 10--17, Sep.~2015.

\bibitem{LK11}
{K.~Lee}, {S.~Kim}, and {B.~Yi}, ``Throughput comparison of random access
  methods for {M2M} service over {LTE} networks,'' in \emph{Proc. IEEE Globecom
  Workshops (GC Wkshps)}.\hskip 1em plus 0.5em minus 0.4em\relax Houston, TX,
  Dec.~2011, pp. 373--377.

\bibitem{IT14}
{M.~Islam}, {A. M.~Taha}, and {S.~Akl}, ``A survey of access management
  techniques in machine type communications,'' \emph{IEEE Comm. Mag.}, vol.~52,
  no.~4, pp. 74--81, Apr.~2014.

\bibitem{OH15}
{C.~Oh}, {D. D.~Hwang}, and {T.~Lee}, ``Joint access control and resource
  allocation for concurrent and massive access of {M2M} devices,'' \emph{IEEE
  Trans. Wireless Commun.}, vol.~14, no.~8, pp. 4182--4192, Aug.~2015.

\bibitem{HH13}
{M.~Hasan} and {E.~Hossain}, ``Random access for machine-to-machine
  communication in {LTE}-advanced networks: {Issues} and approaches,''
  \emph{IEEE Comm. Mag.}, vol.~51, no.~6, pp. 86--93, Jun.~2013.

\bibitem{JKK14}
{H. S.~Jang}, {S. M.~Kim}, {K. S.~Ko}, {J.~Cha}, and {D. K.~Sung}, ``Spatial
  group based random access for {M2M} communications,'' \emph{IEEE Commun.
  Letters}, vol.~18, no.~6, pp. 961--964, Jun.~2014.

\bibitem{ZK15}
{N.~Zhang}, {G.~Kang}, {J.~Wang}, {Y.~Guo}, and {F.~Labeau}, ``Resource
  allocation in a new random access for {M2M} communications,'' \emph{IEEE
  Comm. Letters}, vol.~19, no.~5, pp. 843--846, May.~2015.

\bibitem{LYY16}
C.~Liang, F.~R. Yu, H.~Yao, and Z.~Han, ``Virtual resource allocation in
  information-centric wireless virtual networks,'' \emph{IEEE Trans.\ Veh.\
  Tech.}, 2016, to appear, available online.

\bibitem{KR15}
{D.~Kreutz}, {F. M. V.~Ramos}, {P. E.~Ver\'{\i}ssimo}, {C. E.~Rothenberg},
  {S.~Azodolmolky}, and {S.~Uhlig}, ``Software-defined networking: A
  comprehensive survey,'' \emph{Proc. IEEE}, vol. 103, no.~1, pp. 14--76,
  Jan.~2015.

\bibitem{LYZ15}
{C.~Liang}, {F. R.~Yu}, and {X.~Zhang}, ``Information-centric network function
  virtualization over {5G} mobile wireless networks,'' \emph{IEEE Network},
  vol.~29, no.~3, pp. 68--74, May.~2015.

\bibitem{CYL15}
{Y.~Cai}, {F. R.~Yu}, {C.~Liang}, {B.~Sun}, and {Q.~Yan}, ``Software defined
  device-to-device ({D2D}) communications in virtual wireless networks with
  imperfect network state information ({NSI}),'' \emph{IEEE Trans. Veh. Tech.},
  2015, to appear, available online.

\bibitem{WL16}
{K.~Wang}, {H.~Li}, and {F. R.~Yu}, ``Virtual resource allocation in
  software-defined information-centric cellular networks with device-to-device
  communications and imperfect {CSI},'' \emph{IEEE Trans. Veh. Tech.}, 2016, to
  appear, available online.

\bibitem{BOS14}
{C.~Bernardos}, {A. De La~Oliva}, {P.~Serrano}, {A.~Banchs}, {L.~Contreras},
  {H.~Jin}, and {J.~Zuniga}, ``An architecture for software defined wireless
  networking,'' \emph{IEEE Wireless Commun.}, vol.~21, no.~3, pp. 52--61,
  Jun.~2014.

\bibitem{LM14}
{C.~Luo}, {G.~Min}, {F. R.~Yu}, {Y.~Zhang}, {L. T.~Yang}, and {V.~Leung},
  ``Joint relay scheduling, channel access, and power allocation for green
  cognitive radio communications,'' \emph{IEEE J. Sel. Areas Commun.}, vol.~33,
  no.~5, pp. 922--932, May.~2015.

\bibitem{SS73}
{R. D.~Smallwo} and {E. J.~Sondik}, ``Optimal control of partially observable
  markov processes over a finite horizon,'' \emph{Operations Research},
  vol.~21, pp. 1071--1088, 1973.

\bibitem{LY10}
{C.~Luo}, {F. R.~Yu}, {H.~Ji}, and {V.~Leung}, ``Cross-layer design for {TCP}
  performance improvement in cognitive radio networks,'' \emph{IEEE Trans. Veh.
  Tech.}, vol.~59, no.~5, pp. 2485--2495, June.~2010.

\bibitem{LA04}
{Y.~Lu}, {T. F.~Abdelzaher}, and {A.~Saxena}, ``Design, implementation, and
  evaluation of differentiated caching services,'' \emph{IEEE Trans. Parallel
  Distrib. Syst.}, vol.~15, no.~5, pp. 440--452, May.~2004.

\end{thebibliography}

\end{document}